\documentclass[12pt,twoside,reqno]{amsart}

\usepackage{upref,amssymb}

\theoremstyle{plain}
\newtheorem{theorem}{Theorem}
\newtheorem{lemma}{Lemma}

\theoremstyle{remark}
\newtheorem*{assumption}{Assumption}
\newtheorem{definition}{Definition}
\newtheorem{remark}{Remark}

\newenvironment{acknowledgement}{\emph{Acknowledgement.}}

     {\end{nummer}\end{myremarks}}

\newtheorem{myremarks}[theorem]{Remark}

\newcounter{numcount}
\newcommand{\labelnummer}{\mbox{\normalfont (\roman{numcount})}}%

\makeatletter

\newenvironment{nummer}%
   {\let\curlabelspeicher\@currentlabel%
     \begin{list}{\labelnummer}%
       {\usecounter{numcount}\leftmargin0pt%
         \topsep0.5ex\partopsep2ex\parsep0pt\itemsep0ex\@plus1\p@%
         \labelwidth2.5em\itemindent3.5em\labelsep1em%
       }%
     \let\saveitem\item%
     \def\item{\saveitem%
  \def\@currentlabel{\curlabelspeicher\hskip0.5pt\labelnummer}}%
     \let\savelabel\label%
     \def\label##1{\savelabel{##1}%
       \@bsphack%
         \ifmmode\else%
           \protected@write\@auxout{}%
           {\string\newlabel{##1item}{{\labelnummer}{\thepage}}}%
         \fi%
       \@esphack%
     }%
   }{\end{list}}%

\def\itemref#1{\expandafter\@setref\csname r@#1item\endcsname%
   \@firstoftwo{#1}}%

\newif\ifper\pertrue
\def\per{.}

\newcounter{aucount}
\def\HarvardComma{,}

\def\aufinalstring{,}

\def\edinitstring{}
\def\edfinalstring{ (\editorname),}
\newif\ifedplural
\def\editorname{\ifedplural Eds.\else Ed.\fi}

\def\au#1#2{\textsc{#1 #2}\addtocounter{aucount}{1}}
\def\lau#1#2{\textsc{#1 #2}\aufinalstring\setcounter{aucount}{0}}

\def\ed#1#2{\ifnum\theaucount=0{}\edinitstring\fi\textsc{#1 #2}%
             \addtocounter{aucount}{1}}
\def\led#1#2{\ifnum\theaucount=0{}\edinitstring\edpluralfalse%
                \else\edpluraltrue\fi\textsc{#1 #2}\edfinalstring%
              \setcounter{aucount}{0}}

\def\et{\ifnum\theaucount=1\else\HarvardComma\fi{} and\ }
\def\ti#1{{#1},}

\def\bti{\@ifnextchar[\bbti\bbbti}
\def\bbti[#1]#2{\emph{#2}, #1,}
\def\bbbti#1{\emph{#1},}

\def\z{\@ifnextchar[\zz\zzz}
\def\zz[#1]#2#3#4#5{\perfalse\emph{#2} \textbf{#3} (#5), #4 [#1]\per}
\def\zzz#1#2#3#4{\emph{#1} \textbf{#2} (#4), #3\ifper\per\fi\pertrue}

\def\pub{\@ifstar\pubstar\pubnostar}
\def\pubnostar{\@ifnextchar[\@@pubnostar\@pubnostar}
\def\@@pubnostar[#1]#2#3#4{#1, #2, #3, #4\per}
\def\@pubnostar#1#2#3{#1, #2, #3\ifper\per\fi\pertrue}
\def\pubstar[#1]#2#3#4{\perfalse #2, #3, #4 [#1]\per}

\def\@setauthors{%
   \begingroup
   \trivlist
   \centering\@topsep30\p@\relax
   \advance\@topsep by -\baselineskip
   \item\relax
   \andify\authors
   \def\\{\protect\linebreak}%
  \textsc{\authors}%
  \endtrivlist
   \endgroup
}

\let\isave\i
\def\i{\ifmmode%
            \@ifundefined{comp}%
                  {\mathrm{i}}%
                  {\hspace{0.07em}\mathrm{i}\hspace{0.07em}}%
        \else\isave%
        \fi
}

\newcommand{\pr}{\prime}

\DeclareMathOperator{\e}{e}
\def\d{\mathrm{d}}
\def\Chi{\raisebox{.4ex}{$\chi$}}
\def\Re{\mathop\mathrm{Re}}

\def\tr{\mathop\mathrm{tr}}
\def\tuv{\mathop\mathcal{T}}
\DeclareMathOperator{\supp}{supp}

\def\wlim{\mathop\textup{w-lim}}
\def\slim{\mathop\textup{s-lim}}

\providecommand{\varkappa}{\kappa}

\def\llangle{\langle\mkern-4mu\langle}
\def\rrangle{\rangle\mkern-4mu\rangle}

\def\le{\leqslant}
\def\ge{\geqslant}

\newcommand{\Xicl}{\Xi^{\mathrm{cl}}}

\newcommand{\EE}{\mathbb{E}}
\newcommand{\NN}{\mathbb{N}}
\newcommand{\PP}{\mathbb{P}}
\newcommand{\RR}{\mathbb{R}}

\newcommand{\ZZ}{\mathbb{Z}}
\newcommand{\E}{\mathbb{E}}

\newcommand\bdelta{\boldsymbol{\delta}}

\newcommand\vphi{\varphi}

\newcommand{\cE}{\mathcal{E}}
\newcommand{\cH}{\mathcal{H}}
\newcommand{\cK}{\mathcal{K}}
\newcommand{\cL}{\mathcal{L}}
\newcommand{\cT}{\mathcal{T}}

\newcommand{\cR}{\mathcal{R}}
\newcommand{\cJ}{\mathcal{J}}
\newcommand{\cD}{\mathcal{D}}
\newcommand{\cQ}{\mathcal{Q}}
\newcommand{\cF}{\mathcal{F}}
\newcommand{\cN}{\mathcal{N}}
\newcommand{\cM}{\mathcal{M}}

\newcommand{\cS}{\mathcal{S}}

\newcommand{\dX}{\dot{X}_{1}}   

\newcommand{\omL}{\mkern2mu\raisebox{1.5pt}{$\scriptscriptstyle\odot_{L}^ 
{\phantom{J}}$}\mkern1mu}
\newcommand{\omR}{\mkern2mu\raisebox{1.5pt}{$\scriptscriptstyle\odot_{R}^ 
{\phantom{J}}$}\mkern2mu}

\newcommand{\tnorm}[2][]{\ifx @#1@\def\myleft{\left}%
          \def\myright{\right}%
     \else\def\myleft{\csname#1l\endcsname}%
          \def\myright{\csname#1r\endcsname}%
          \ifx\myleft\normall\def\myleft{\relax}\def\myright\relax\fi%
     \fi%
     \myleft|\!\myleft|\!\myleft| #2 \myright|\!\myright|\!\myright|}

\newcommand{\norm}[2][]{\ifx @#1@\def\myleft{\left}%
          \def\myright{\right}%
     \else\def\myleft{\csname#1l\endcsname}%
          \def\myright{\csname#1r\endcsname}%
          \ifx\myleft\normall\def\myleft{\relax}\def\myright\relax\fi%
     \fi%
     \myleft\Vert #2 \myright\Vert}

\newcommand{\abs}[1]{\left\lvert #1 \right\rvert}
\newcommand{\set}[1]{\left\{ #1 \right\}}

\newcounter{pacount} 

\newcommand{\pa}[2][]{\addtocounter{pacount}{1}%
  \ifx @#1@\expandafter\def\csname myleftp\thepacount\endcsname{\left}%
          \expandafter\def\csname myrightp\thepacount\endcsname{\right}%
     \else\expandafter\def\csname myleftp\thepacount\endcsname{\csname#1l\endcsname}%
          \expandafter\def\csname myrightp\thepacount\endcsname{\csname#1r\endcsname}%
          \ifx\csname myleftp\thepacount\endcsname\normall\expandafter\def\csname myleftp\thepacount\endcsname{\relax}\expandafter\def\csname myrightp\thepacount\endcsname{\relax}\fi%
     \fi%
     \csname myleftp\thepacount\endcsname( #2 \csname myrightp\thepacount\endcsname)\addtocounter{pacount}{-1}}

\newcommand{\hnorm}[1]{\left\{ \!\left\{ #1\right\}\! \right\}}

\newcommand\beq{\begin{equation}}
\newcommand\eeq{\end{equation}}

\newcommand{\eq}[1]{\eqref{#1}}

\numberwithin{equation}{section}
\makeatother

\begin{document}


\title{The conductivity measure for the Anderson model}

\author[Abel Klein]{Abel Klein$^{1}\!$}
\address{ University of
   California, Irvine, Department of Mathematics, CA 92697-3875, USA}
\email{aklein@math.uci.edu}{}

\author[Peter M\"uller]{Peter M\"uller}
\address{Institut f\"ur Theoretische Physik, Friedrich-Hund-Platz~1,
   Georg-August-Universit\"at, 37077 G\"ottingen, Germany}
\email{peter.mueller@physik.uni-goe.de}

\thanks{$^{1}$Supported in part by NSF Grant DMS-0457474.}


\dedicatory{Dedicated to Leonid A.\ Pastur on the occasion of his 70th
   birthday}

\begin{abstract}
  We study the ac-conductivity in linear response theory for the
  Anderson tight-binding model.  We define the electrical
  ac-conductivity and calculate the linear-response current at zero
  temperature for arbitrary Fermi energy. In particular, the Fermi
  energy may lie in a spectral region where extended states are
  believed to exist.
 \end{abstract}

\maketitle

%

\section{Introduction}

In this article we study the ac-conductivity in linear response theory for the Anderson tight-binding model.  We  define  the electrical
ac-conductivity and calculate the  linear-response current   at temperature $T=0$  for arbitrary Fermi energy $\mu$.

 At temperature $T=0$, if the
 Fermi energy $\mu$ is either in the region of localization or
outside the spectrum of the random Schr\"odinger operator, this was already
done in \cite{KLM} by a careful mathematical analysis of the
ac-conductivity in linear response theory, following the approach of
\cite{BGKS}, and the introduction of a new concept, the conductivity
measure.   This approach can be easily  extended to the nonzero temperature case, $T>0$,
with $\mu$ (here the chemical potential) arbitrary.     The conductivity
measure $ \Sigma_{\mu}^{T}(\d \nu)$, with  $\nu$  the frequency of the applied electric field, is a finite positive even Borel measure on the real line. If $ \Sigma_{\mu}^{T}(\d
\nu)$ was known to be an absolutely continuous measure, the {in-phase}
or {active} conductivity $ \Re \sigma_{\mu}^T(\nu)$ would then be
well-defined as its density. The conductivity measure $
\Sigma_{\mu}^{T}(\d \nu)$ is thus an analogous concept to the density
of states measure $\mathcal{N}(\d E)$, whose formal density is the
density of states $n(E)$.   Given a   spatially homogeneous, time-dependent
electric field $\mathbf{E}(t)$, the in-phase linear-response
current at time $t$, $ J_{\mathrm{lin}}^{\mathrm{in}}(t;  \mu,T,\mathbf{E})$, has a simple expression in terms of this conductivity measure:
\begin{equation}\label{JinInt}
  J_{\mathrm{lin}}^{\mathrm{in}}(t; \mu,T,\mathbf{E}) =
  \int_{\mathbb{R}}\!\Sigma_{\mu}^{T}(\d\nu)\; \e^{\i\nu t}
  \widehat{\mathbf{E}}(\nu).
\end{equation}

This procedure is conjectured to break down at $T=0$ for, say, Fermi energies
$\mu$ in the region of extended states.  In this case there has been no
suitable derivation of the in-phase linear-response current.  In this paper we
define the conductivity measure $ \Sigma_{\mu}^{0}(\d \nu)$ and the in-phase
linear-response current for \emph{arbitrary} Fermi energy $\mu$.  We give an
explicit expression for $ \Sigma_{\mu}^{0}( \d\nu)$, and justify the definition
by proving that 
\beq \Sigma_{\mu}^{0}(\d\nu)= \lim_{T\downarrow 0}
\Sigma_{\mu}^{T}(\d\nu)\quad \text{weakly for Lebesgue-a.e.\ $\mu\in\RR$ }.
\eeq
 The in-phase linear-response current is then defined by \eqref{JinInt},
and justified by
 \beq J_{\mathrm{lin}}^{\mathrm{in}}(t; \mu,0,\mathbf{E})=
\lim_{T\downarrow 0} J_{\mathrm{lin}}^{\mathrm{in}}(t; \mu,T,\mathbf{E}) \quad
\text{for Lebesgue-a.e.\ $\mu\in\RR$ }.  \eeq

\begin{acknowledgement} 
  This paper originated from discussions with {Leonid A.\ Pastur}, to
  whom this paper is dedicated on the occasion of his 70th birthday.
  Pastur is a founding father of the theory of random Schr\"odinger
  operators; of particular relevance to this paper is his work on the
  electrical conductivity, e.g.,
  \cite{BePa70,Pas71,Pas73,LiGr88,KP,Pas99a,Pas99b,KiLe03}.  The
  authors also thank Olivier Lenoble for many discussions.
\end{acknowledgement}

\section{Definitions and Results}
The Anderson tight-binding model is described by the random
Schr\"odinger operator $H$, a measurable map $\omega \mapsto
H_{\omega}$ from a probability space $(\Omega,\mathbb{P})$ (with
expectation $\mathbb{E}$) to bounded self-adjoint operators on
$\ell^2(\ZZ^d)$, given by
\begin{equation}
   \label{AND}
   H_\omega := - \Delta + V_\omega .
\end{equation}
Here $\Delta$ is the centered discrete  Laplacian,
\begin{equation}
  (\Delta \varphi)(x):= -  \sum_{{y\in\ZZ^d; \, |x-y|=1}} \varphi(y)  
\qquad
  \text{for} \quad   \varphi\in\ell^2(\ZZ^d),
\end{equation}
and the random potential $V$ consists of independent, identically
distributed random variables $\{V(x) ; x \in \ZZ^d\}$ on
$(\Omega,\mathbb{P})$, such that the common single site probability
distribution has a bounded density $\rho$ with compact support.

The \emph{Anderson Hamiltonian} $H$ given by \eqref{AND} is
$\mathbb{Z}^d$-ergodic, and hence its spectrum, as well as its
spectral components in the Lebesgue decomposition, are given by
non-random sets $\PP$-almost surely \cite{KiMa82, CaLa90, PaFi92}.   This non-random spectrum will be denoted by $\mathfrak{S}$,  with $\mathfrak{S}_{\varkappa}$, $\varkappa=$ pp, ac, sc, denoting its non-random spectral components.

We now outline the derivation of electrical ac-conductivities within linear response theory for the Anderson model. We refer to
\cite{BGKS} and \cite{KLM} for mathematical details, generalizations
and proofs.

At the reference time $t=-\infty$, the system is assumed to be in
thermal equilibrium at absolute temperature $T\ge 0$ and  chemical
potential $\mu\in\RR$. On the single-particle level, this equilibrium
state is given by the random operator $f_{\mu}^{T}(H)$, where
\begin{equation}
  \label{fermi}
  f_{\mu}^{T}(E) := 
  \begin{cases}
    \pa[Big]{ \e^{ \frac{E-\mu}T } +1}^{-1} & \text{if} \quad  T>0  \smallskip \\
    \Chi_{]-\infty, \mu]}(E) &\text{if} \quad  T=0
  \end{cases}
  \end{equation}
stands for the Fermi function. By $\Chi_{B}$ we denote the indicator
function of the  set $B$. A spatially homogeneous, time-dependent
electric field $\mathbf{E}(t)$ is then introduced adiabatically:
Starting at time $t= -\infty$, we switch on the (adiabatic) electric
field $\mathbf{E}_{\eta}(t):= \e^{\eta t}\mathbf{E}(t)$ with $\eta
>0$, and then let $\eta \to 0$. 

On account of isotropy we assume
without restriction that the electric field is pointing in the
$x_{1}$-direction: $\mathbf{E}(t)=\mathcal{E}(t) \widehat{x}_{1} $,
where $\mathcal{E}(t)$ is the (real-valued) amplitude of the electric
field, and $\widehat{x}_{1}$ is the unit vector in the
$x_{1}$-direction.  Our precise requirements for the real-valued,
time-dependent amplitude $\cE(t)$ are stated in the following
assumption, which we assume valid from now on.

\begin{assumption}[$\mathsf{E}$]  The time-dependent amplitude $\cE(t)$
of the electric field is of the form
   \begin{equation}
     \cE(t) =  \int_{\RR}\!\d \nu \; \e^{\i\nu t}\widehat{\cE}(\nu) ,
   \end{equation}
  where  $\widehat{\cE} \in C(\RR)\cap \mathrm{L}^1(\RR)$  with
   $\widehat{\cE}(\nu)=\overline{ \widehat{\cE}(-\nu)}$.
\end{assumption}

For each $\eta>0$ this procedure results in a time-dependent random Hamiltonian
\begin{equation}
   H_{\omega}({\eta, t}) := G(\eta,t) H_{\omega} G(\eta,t)^{*}, \quad \text{with} \quad
   G(\eta,t) := \e^{\i X_{1} \int_{-\infty}^{t}\d s\,
     \e^{\eta s} \cE(s)},
\end{equation}
where $X_{1}$ stands for the operator of multiplication by the first
coordinate of the electron's position.  $ H_{\omega}({\eta, t})$ is,
of course, gauge equivalent to $H_{\omega} + \e^{\eta t} \cE(t)
X_{1}$.  At time $t$, the state of the system is described by the
random operator $\varrho_{\omega}(\eta,t)$, the solution to
the Liouville equation
\begin{equation}\label{Liouvilleeq}
   \left\{
     \begin{array}{l}
     \mbox{$\i$} \partial_t \varrho_{\omega}(\eta,t) =
     [H_{\omega}(\eta,t),\varrho_{\omega}(\eta,t)]
       \\
       \displaystyle\lim_{t \to  -\infty}\varrho_{\omega}(\eta,t)=
       f_{\mu}^{T}(H_{\omega})
   \end{array}
   \right.  .
\end{equation}

The adiabatic electric field generates a time-dependent electric
current. Thanks to reflection covariance  in all but the
first direction, the current is also oriented along
the first coordinate axis. Its amplitude is
\begin{equation}
   \label{curdef}
   J_{\eta}(t;\mu,T,\mathcal{E}) = -  \cT \bigl( \varrho_{\omega} 
(\eta,t)
   \dot{X}_{1}(t)\bigr),
\end{equation}
where $\cT$ is the trace per unit volume (see \eqref{tuv1} and
\eqref{tuv} in Appendix~\ref{sec:Kubo}) and $ \dX$ is the first
component of the velocity operator:
\begin{equation}
   \dX := \i [H_{\omega}, X_{1}] = \i [-\Delta, X_{1}].
\end{equation}
Note that we are using the Schr\"odinger picture in \eqref{curdef}.
The time dependence of the velocity operator $\dot{X}_{1}(t) :=
G(\eta,t)\dX G(\eta,t)^{*}$ there results from our particular gauge.
Finally, the \emph{adiabatic linear-response current} is defined as
\begin{equation}
   \label{lin-cur}
   J_{\eta,\mathrm{lin}}(t;\mu,T,\cE) := \frac{\d}{\d\alpha}\,
   J_{\eta }(t;\mu,T,\alpha\cE)\big|_{\alpha=0}.
\end{equation}

The detailed analysis in \cite{BGKS} shows that one can give a
mathematical meaning to the formal procedure leading to
\eqref{lin-cur}, for fixed temperature $T\ge 0$ and chemical potential
$\mu \in \RR$, if the corresponding thermal equilibrium random
operator $ f_{\mu}^{T}(H)$ satisfies the condition
\begin{equation} 
  \label{assumpIntro} 
  \E\big\{ \big\| X_{1}\,
      f_{\mu}^{T}(H_{\omega}) \delta_0\big\|^2 \big\} < \infty ,
\end{equation}
where $\set{\delta_a}_{a \in \ZZ^d}$ is the canonical orthonormal basis in $\ell^2(\ZZ^d)$: $\delta_a(x)=1 $ if $x=a$ and $\delta_a(x)=0$ otherwise.  (This is the condition originally identified in \cite{BESB}.) 

The derivation of a Kubo formula for the ac-conductivity
\cite{BESB,SBB,BGKS} requires normed spaces of measurable covariant
operators. The required mathematical framework is described in
Appendix~\ref{sec:Kubo}; here we will be somewhat informal.  $\cK_{2}$
is the Hilbert space of measurable covariant operators $A$ on
$\ell^2(\ZZ^d)$, i.e., measurable, covariant maps $\omega \mapsto
A_{\omega}$ from the probability space $(\Omega,\mathbb{P})$ to
operators on $\ell^2(\ZZ^d)$, with inner product
\begin{equation}
  \label{scalarproduct}
  \llangle A, B\rrangle :=  \ \EE \bigl\{\langle
  A_{\omega}\delta_{0}, B_{\omega}  
  \delta_{0}\rangle
  \bigr\} =  \cT \set{A_{\omega}^{*} B_{\omega} }
\end{equation}
and norm $ \tnorm{{A}}_2:= \sqrt{ \llangle A, A\rrangle}$. Here $\cT$,
given by $\cT(A) := \EE \{\langle\delta_{0}, A_{\omega}
\delta_{0}\rangle \}$, is the trace per unit volume.  The Liouvillian
$\cL$ is the (bounded in the case of the Anderson model) self-adjoint
operator on $\cK_2$ given by the commutator with $H$: 
\beq \pa{\cL
  A}_\omega:= [H_\omega, A_\omega].  
\eeq 
We also introduce operators
$\cH_L$ and $ \cH_R$ on $\cK_2$ given by left and right multiplication
by $H$: 
\beq \pa{\cH_L A}_\omega:= H_\omega A_\omega \quad \text{and}
\pa{\cH_R A}_\omega:= A_\omega H_\omega .  
\eeq 
Note that $\cH_L$ and
$ \cH_R$ are commuting, bounded (for the Anderson Hamiltonian),
self-adjoint operators on $\cK_2$, anti-unitarily equivalent (see
\eq{JHJ}), and $\cL =\cH_L - \cH_R$.  It follows from the Wegner
estimate for the Anderson Hamiltonian that in this case the operators
$\cH_L$ and $ \cH_R$ have purely absolutely continuous spectrum (see
Lemma~\ref{lemHLabscts} in Section~\ref{secProofs}).  For each $T \ge
0$ and $\mu \in \RR$ we consider the bounded self-adjoint operator
$\cF_\mu^T$ in $\cK_2$ given by
 \begin{equation} \label{defFmuT}
  \cF_\mu^T := f_{\mu}^{T}(\cH_{L}) - f_{\mu}^{T}(\cH_{R})   , \quad \text{i.e.}, \quad \pa[big]{\cF_\mu^T A}_\omega= [f_\mu^T(H_\omega),A_\omega]. 
  \end{equation} 
In this setting the key condition \eq{assumpIntro} may be rewritten as
\begin{equation}\label{defY}
Y_{\mu}^{T} := \i [X_{1}, f_{\mu}^{T}(H)] \in \cK_{2}.
\end{equation}

Note that condition \eqref{defY} is always true for $T>0$ with arbitrary $\mu \in \RR$,
since in this case $f_{\mu}^{T}(H)= g(H)$ for some $g \in \mathcal{S}(\RR^d)$
(cf.\ \cite[Remark~5.2(iii)]{BGKS}).  We set
\begin{align}
  \Xi_0 := 
  \set{\mu \in \RR; \quad Y_{\mu}^{0}  \in \cK_{2}}.
\end{align}
For the same reason as when $T>0$, we have  $\mu
\in \Xi_0 $ if either $\mu \notin \mathfrak{S}$ or $\mu$ is the left edge of a
spectral gap for $H$.  Moreover, letting $\Xicl$ denote the region of complete localization,  defined as the region of
validity of the multiscale analysis, or equivalently, of the fractional moment
method, we have  (cf. \cite{AiGr98,GKjsp})
\beq \label{XiclXi0}
\Xicl \subset \Xi_0 .
\eeq

  A precise definition of the region of complete localization is given in Appendix~\ref{Xicl}. Note that we
included the complement of the spectrum $\mathfrak{S}$ in $\Xicl$ for
convenience, and that  $\Xicl$ is an open set by its definition.  Note also that
   for $\mu \in \Xicl$ the Fermi projection $f_{\mu}^{0}(H)$ satisfies a
  much stronger condition than \eq{assumpIntro}, namely exponential decay of
  its kernel \cite[Theorem~2]{AiGr98} (see \eq {fermidecay2}).
 Conversely, fast enough polynomial decay
  of the kernel of the Fermi projection for all energies in an interval
  implies complete localization in the interval \cite[Theorem~3]{GKjsp}.

If $ Y_{\mu}^{T} \in \cK_{2}$, we  proceed as  in \cite{KLM}, with a slight variation to include also the case when   $T>0$. 
An inspection of the proof of   \cite[Thm.~5.9]{BGKS} shows that  the adiabatic linear-response current \eqref{lin-cur} is  well
defined for every time $t\in\RR$, and given by
\begin{equation}
     \label{lin-cur2}
     J_{\eta,\mathrm{lin}}(t;\mu,T,\cE) = \tuv \left\{
       \int_{-\infty}^{t}\!\d s\; \e^{\eta s} \cE(s) \dX
       \e^{-\i (t-s) \cL} Y_{\mu}^{T}
     \right\}.
   \end{equation}

It is convenient to rewrite \eq{lin-cur2} in terms of the
 conductivity measure $ \Sigma_{\mu}^{T}$, which we now introduce  if either $T>0$ or $\mu \in \Xi_0 $.

\begin{definition}\label{defSigma} If either $T>0$ or $\mu \in \Xi_0 $,
  the  \emph{(ac-)conductivity measure} ($x_{1}$-$x_{1}$ component) at temperature $T$ and chemical potential $\mu$
  is defined by
  \begin{equation}
     \label{kubo-expr}
     \Sigma_{\mu}^{T}(B) := \pi \llangle \dot{X}_{1},
     \Chi_{B}(\mathcal{L}) Y_{\mu}^{T} \rrangle  \quad \text{for all Borel sets $B\subset\mathbb{R}$}.
\end{equation}
\end{definition}

This definition is justified by the following theorem, whose proof, as
the proofs of all other results in this section, is postponed to
Section~\ref{secProofs}.  $\mathcal{M}(\RR)$ will denote the vector
space of complex Borel measures on $\RR$, with $\mathcal{M}_+(\RR)$
being the cone of finite positive Borel measures, and with
$\mathcal{M}_+^{(\e)}(\RR)$ the finite positive even Borel measures.
We recall that $\mathcal{M}(\RR)=C_0(\RR)^*$, where $C_0(\RR)$ denotes
the Banach space of complex-valued continuous functions on $\RR$
vanishing at infinity with the sup norm.  We will use two locally
convex topologies on $\mathcal{M}(\RR)$. The first is the weak$^*$
topology, defined by the linear functionals $\set{\Gamma \in
  \mathcal{M}(\RR) \mapsto \Gamma(g); \; g \in C_0(\RR)}$. (By
$\Gamma(g) := \int_{\RR} \Gamma(\d s)\, g(s)$ we denote the integral of a
function $g$ with respect to a measure $\Gamma$.)  The second is the
one defined by the similarly defined linear functionals where $g$ is
any bounded measurable function on $\RR$.  `Weak' will refer to the
weak$^*$ topology and `strong' to the other topology.  We will write
$\wlim$ and $\slim$ to denote the respective limits.

\begin{theorem} \label{thmSigma}
\begin{nummer}
 \item If either $T>0$ or $\mu \in \Xi_0 $, the conductivity measure $ \Sigma_{\mu}^{T}$ is a  finite positive even Borel measure on the real line, i.e., $ \Sigma_{\mu}^{T}\in \mathcal{M}_+^{(\e)}(\RR)$, such that
\beq \label{SigmaR}
 \Sigma_{\mu}^{T}(\RR)= - \pi \mathbb{E} \bigl\{ \langle \delta_{\widehat{x}_{1}} +
     \delta_{-\widehat{x}_{1}}, f_{\mu}^{T}(H)\delta_{0}\rangle \bigr 
\} \le \sqrt{2}\, \pi .
\eeq

\item \label{sameKLM}  For every  $\mu \in \Xi_0 $ we have
\beq \label{SigmaKLM}
 \Sigma_{\mu}^{0}(B) = \pi \llangle Y_{\mu}^{0},
     \Chi_{B}(\mathcal{L})\pa{- \cL}   \cF_\mu^0 Y_{\mu}^{0} \rrangle  \quad \text{for all Borel sets $B\subset\mathbb{R}$}.
\eeq

\item \label{T>0cont}  The
     map $]0,\infty[ \ni T \mapsto \Sigma_{\mu}^{T} \in \mathcal{M}_+^{(\e)}(\RR)$ is  
strongly continuous  for every $\mu\in\RR$.

\item \label{slimSig}For every $\mu \in \Xi_0$ we have
\beq  \slim_{T\downarrow 0} \Sigma_{\mu}^{T} = \Sigma_{\mu}^{0}.
\eeq

\item  \label{slimXicl}
If   $\mu \in \Xicl$ we also have  $ \lim_{T\downarrow 0} Y_{\mu}^{T}=Y_{\mu}^{0}$ in $\cK_2$.
\end{nummer}
\end{theorem}

\begin{remark} 
\begin{nummer}
\item Theorem~\ref{sameKLM}    shows that for $T=0$ and $\mu \in \Xi_0 $ the  conductivity measure $ \Sigma_{\mu}^{0}$ defined by \eq{kubo-expr}  coincides with the one given in  \cite[Definition~3.3]{KLM}.

 \item  If the Fermi energy $\mu$ is above or below the almost-sure spectrum
   $\mathfrak{S}$ of $H$, we have $Y_{\mu}^{0}=0$, and hence also
   $\Sigma_{\mu}^{0} =0$. If $]a,b[$ is a spectral gap, we clearly have
   $Y_{\mu}^{0}=Y_{a}^{0}$, and hence $\Sigma_{\mu}^{0}= \Sigma_{a}^{0}
   $, for all $\mu \in ]a,b[$. Moreover, it is shown in
   \cite[Proposition~3.7]{KLM} that the measure $\Sigma_{\mu}^{0}$
   can be expressed in terms of a measure $ \Psi_\mu$ on $\RR^2$,
   supported by the set $\mathbb{S}_\mu$ given in
   \cite[Eq.~(3.41)]{KLM}. Since $ \Psi_\mu$ depends on $\mu$ only
   through $Y_{\mu}^{0}$, we have $ \Psi_\mu=\Psi_a$ for all $\mu \in
 ]a,b[$, and hence $ \Psi_a$ is supported by the set
   \begin{equation}
     \bigcap_{\mu \in [a,b[} \mathbb{S}_\mu  =   \bigl\{]-\infty,a]
     \times [b,\infty[\bigr\}
     \cup \bigl\{ [b,\infty[\times ]-\infty,a]\bigr\} .
        \end{equation}
    It then follows from  \cite[Eq.~(3.40)]{KLM} that for all $\mu \in
    [a,b[$ we have
   \begin{equation}
   \Sigma^0_{\mu} ([-\nu,\nu]) = \Sigma^0_a ([-\nu,\nu]) = 0 \quad
   \text{for all $\nu \in ]0, b-a[$. }
    \end{equation}
    
    \item  \label{remdirect} If $\mu \in \Xi_{0}$,  as
   shown in \cite{Nak02,BGKS}, the direct-current conductivity
   vanishes  at zero temperature:
   \begin{equation}
     \sigma^{0}_{\mu,\mathrm{dc}} := \lim_{\eta\downarrow 0} \; 
     \Big\langle\mkern-6mu\Big\langle \dX,
     \frac{1}{\i\cL +\eta} \,Y_{\mu}^{0} \Big\rangle\mkern-6mu\Big\rangle = 0 .
   \end{equation}
\item   For  $\mu \in \Xicl$,   the region of complete  localization,  the Mott-type
   bound
   \begin{equation}
     \limsup_{\nu\downarrow 0} \frac{\tfrac{1}{\nu}
       \Sigma_{\mu}^{0}([0,\nu])}{\nu^{2} \Bigl(\log
       \tfrac{1}{\nu}\Bigr)^{d+2}} \le  \mathrm{constant}
   \end{equation}
   for the ac-conductivity measure was established in \cite{KLM}.

    \end{nummer}
    \end{remark}\medskip

We  may now rewrite \eq{lin-cur2} in terms of the conductivity measure as follows.  If either $T>0$ or $\mu \in \Xi_0 $,  the same argument leading to  \cite[Eq.~(3.30) and Theorem~3.4]{KLM} gives
\begin{equation}
  \label{lin-cur25}
  J_{\eta,\mathrm{lin}}(t;\mu,T,\cE) =  \e^{\eta t}  \int_{\RR}
  \!\d\nu \,\e^{\i\nu t} 
  \sigma_{\mu}^{T}(\eta,\nu) \,
  \widehat{\cE}(\nu),
\end{equation}
where $ \sigma_{\mu}^{T}(\eta,\cdot)$ is the Stieltjes transform of
the conductivity measure $\Sigma_{\mu}^{T}$:
\begin{equation}
  \sigma_{\mu}^{T}(\eta,\nu) := - \frac{\i}{\pi}
  \int_{\mathbb{R}}\!
  \Sigma_{\mu}^{T}(\d\lambda)\; \frac{1}{\lambda +\nu +\i\eta}.
\end{equation}
The adiabatic in-phase linear-response current is now  defined by
\begin{equation}
  J_{\eta,\mathrm{lin}}^{\mathrm{in}}(t;\mu,T,\cE) := \e^{\eta t}
  \int_{\RR} \!\d\nu \,\e^{\i\nu t}
  \bigl( \Re\sigma_{\mu}^{T}(\eta,\nu)\bigr) \,
  \widehat{\cE}(\nu).
\end{equation}
Turning off the adiabatic switching, we obtain a simple expression for
the in-phase linear-response current in terms of the conductivity
measure, as in \cite[Corollary~3.5]{KLM}, given by
\begin{equation}\label{Jin}
  J_{\mathrm{lin}}^{\mathrm{in}}(t; 
  \mu,T,\cE):=   \lim_{\eta\downarrow0} J_{\eta,\mathrm{lin}}^{\mathrm{in}}(t; 
  \mu,T,\cE) =
  \int_{\mathbb{R}}\!\Sigma_{\mu}^{T}(\d\nu)\; \e^{\i\nu t}
  \widehat{\mathcal{E}}(\nu).
  \end{equation}
This gives a derivation of the in-phase linear-response current
\eqref{JinInt}, and \eq{Jin} is valid as long as either $T>0$ or $\mu \in
\Xi_0 $.  Moreover, it follows from   \eq{Jin} and Theorem~\ref{slimSig} that
\beq \label{Jin0}
 J_{\mathrm{lin}}^{\mathrm{in}}(t; 
  \mu,0,\cE)=   \lim_{T \downarrow 0}  J_{\mathrm{lin}}^{\mathrm{in}}(t; 
  \mu,T,\cE) \quad \text{for all} \;\; \mu \in \Xi_0.
\eeq

We have so far  constructed  the conductivity measure and the  in-phase linear-response current at $T=0$ if  $\mu \in \Xi_{0}$. But what if, say, there is absolutely continuous spectrum and  $\mu \in \mathfrak{S}_{ac}$?  In this case there is no reason to expect $ \mu \in \Xi_0$.  In view of Remark~\ref{remdirect} we  conjecture   that $ \mu \notin \Xi_0$ for most $\mu \in \mathfrak{S}_{ac}$.

In  this article we show that  the conductivity measure at zero
temperature can be constructed for \emph{arbitrary} Fermi energy $\mu$ in a
physically sensible way as the weak limit of the finite-temperature
conductivity measures as $T\downarrow 0$, with the corresponding  in-phase linear-response current  given by \eq{Jin0}.

To motivate  our construction, we take $T>0$ and decompose $\Sigma_{\mu}^{T}$ as
\beq
\Sigma_{\mu}^{T}= \Sigma_{\mu}^{T}\pa{\set{0}}\bdelta_{0} + \pa{\Sigma_{\mu}^{T} -   \Sigma_{\mu}^{T}\pa{\set{0}}\bdelta_{0}},
\eeq
where the Dirac measure $\bdelta_{0}$ is the Borel measure on $ \RR$
concentrated at $0$ with total measure one.  The details of this
decomposition, presented in the following theorem, will lead to a
natural definition of $\Sigma_{\mu}^{0}$ for arbitrary $\mu$.  We
recall that the Anderson model satisfies the Wegner estimate
\cite{Weg81}, and hence the density of states measure $\cN \in
\cM_+(\RR)$, defined by
\beq\label{defN}
\cN(B): = \cT \pa{\Chi_B(H) }= \EE \{\langle\delta_{0}, \Chi_B(H_\omega) \delta_{0}\rangle \}
\quad \text{for all Borel sets $B\subset\mathbb{R}$},
\eeq
 supported by the spectrum  $\mathfrak{S}$ of $H$, is  absolutely continuous with density
 $n$ satisfying  $\norm{n}_{\infty} \le  \norm{\rho}_{\infty}$.

 We will use the following convention: If $\Gamma \in \cM_+(\RR)$ is absolutely continuous and supported by the closed set $F \subset \RR$, we always assume that 
 its  density $\gamma$ is also supported by $F$.

 We set
\begin{align}
\cQ_{0}:= \Chi_{\set{0}}(\cL) \quad \text{and} \quad \cQ_{\perp}:=I - \cQ_{0},
\end{align}
the orthogonal projections onto the kernel of $\cL$ in $\cK_{2}$ and its orthogonal complement.  Note that $\cQ_0$ and $\cQ_{\perp}$ commute with $\cH_L$ and $\cH_R$, and we have
\beq\label{QH=QH}
g( \cH_L )\cQ_0 =g (\cH_R) \cQ_0 \quad \text{for all bounded Borel functions $g$}.
\eeq  
For each $T \ge 0$ and $\mu \in \RR$, the bounded self-adjoint operator $\cF_\mu^T$, defined in \eq{defFmuT},  satisfies
\beq
 \label{FQ}
 \cQ_{0}  \cF_\mu^T =   \cF_\mu^T \cQ_{0}= 0 \quad \text{and} \quad  
\cF_\mu^T=  \cF_\mu^T  \cQ_\perp =   \cQ_\perp \cF_\mu^T.
\eeq
We let $\cL_\perp^{-1}$ denote the pseudo-inverse to $\cL$, that is,
\beq
\cL_\perp^{-1} := g(\cL) \quad \text{with
$g(t) := \frac 1 t$ if $t\not= 0$ and $g(0)=0$}.
\eeq In particular, 
\beq
 \cL_\perp^{-1}\cL = \cQ_\perp.  \label{HQL}
\eeq 
Moreover, we have $ - \cL \cF_\mu^T \ge 0$ and
\beq \label{LF2}
- \cL_\perp^{-1}\cF_\mu^T=F_\mu^T(\cH_L,\cH_R),
\eeq
where
\beq  \label{LF22}
F_\mu^T(\lambda_1,\lambda_2):=\begin{cases}- \frac {f_\mu^T(\lambda_1) -f_\mu^T(\lambda_2)}{\lambda_1-\lambda_2}= \abs{ \frac {f_\mu^T(\lambda_1) -f_\mu^T(\lambda_2)}{\lambda_1-\lambda_2}} & \text{if} \quad \lambda_1\not= \lambda_2\\
0 & \text{otherwise}
\end{cases}.
\eeq
We write $\mathcal{D}(\mathcal{A})$ for the domain of an unbounded operator
$\mathcal{A}$ in $\cK_{2}$.

\begin{theorem}
  \label{main0}
  \begin{nummer}
  \item\label{main0i} 
    Let
    \begin{equation}
      \label{defPsi}
      \Psi(B) :=\pi  \llangle \dX,\cQ_0
      \,\Chi_{B}(\cH_L)\dX\rrangle  \quad \text{for all Borel sets
        $B\subset\mathbb{R}$}. 
    \end{equation}
    Then $\Psi \in \cM_+(\RR)$ is absolutely continuous with respect
    to the density of states measure $\cN$, and its density with
    respect to Lebesgue measure, $\psi$, satisfies $\psi(E) \le 4\pi
    n(E)\le 4\pi \norm{\rho}_{\infty}$ for Lebesgue-a.e.\ $E\in\RR$.
    Moreover, we have   $\supp \Psi\subset \overline{ \RR \setminus \Xi_{0}}\subset  \RR \setminus \Xicl$.
    
      \item\label{finiteGamma} 
    For each $T\ge 0$ and $\mu \in \RR$ we
    have $\dX \in \cD\pa[Big]{\pa[normal]{ -
        \cL_\perp^{-1}\cF_\mu^T}^{\frac 1 2}}$.  Setting
    \beq  
      \label{defGamma1}
      \Gamma_{\mu}^{T}(B) := \pi  \llangle \pa[big]{ -
        \cL_\perp^{-1}\cF_\mu^T}^{\frac 1 2} \dX, \Chi_{B}(\cL)\pa[big]{
        - \cL_\perp^{-1}\cF_\mu^T}^{\frac 1 2} \dX\rrangle   
    \eeq 
    for all Borel sets $B\subset\mathbb{R}$, we have $\Gamma_{\mu}^{T}
    \in \cM_+^{(\e)}(\RR)$ with $ \Gamma_{\mu}^{T}(\set{0})=0$.  
  \item 
    If either $T>0$ or $\mu \in \Xi_0 $, we have $ \cF_{\mu}^{T}
    \dX\in \cD(\cL_\perp^{-1})$ and
    \beq \label{defGamma12}
      \Gamma_{\mu}^{T}(B) = \pi \llangle \dX,
      \Chi_{B}(\cL)\pa[big]{-\cL_\perp^{-1} \cF_{\mu}^{T}} \dX\rrangle
      \quad \text{for all Borel sets $B\subset\mathbb{R}$}.  
    \eeq   
  \item\label{main03} 
    For all $T>0$ and $\mu \in \RR$ we have
    \begin{align}\label{Sigma0}
      \Sigma_{\mu}^{T}\pa{\set{0}} &= \Psi\pa[big]{
        \pa[normal]{-f_{\mu}^{T}}^{\pr}},\\
      \Sigma_{\mu}^{T}\pa{B\setminus \set{0}} & =
      \Gamma_{\mu}^{T}(B)  \quad \text{for all Borel sets
        $B\subset\mathbb{R}$},      
    \end{align}
    yielding the following decomposition of the conductivity measure
    into mutually singular measures: 
    \beq \label{Sigmadecomp}
      \Sigma_{\mu}^{T} = \Psi\pa[big]{ \pa[normal]{-
          f_{\mu}^{T}}^{\pr}}\bdelta_{0} + \Gamma_{\mu}^{T}.  
    \eeq  
  \item \label{SigmaGamma}
    For all $\mu \in\Xi_0$ we have
    \beq \label{SigmaGamma0}
      \Sigma_{\mu}^{0} = \Gamma_{\mu}^{0}. 
    \eeq
  \end{nummer}
\end{theorem}

\begin{remark} \label{remfTlim} 
  On account of Theorem~\ref{main0i} we
  assume without loss of generality that $\psi(\mu)=0$ for \emph{all} $\mu
  \in \Xi_0$.
\end{remark}

\begin{remark} The measure $\Gamma_{\mu}^{T}$  given in \eq{defGamma1} can be expressed in terms of  the velocity-velocity correlation measure $\Phi \in \cM_+(\RR^2)$, defined by (cf. \cite[Eq.~(3.46)]{KLM}) 
 \beq \label{vel-vel}
\Phi(C):=  \llangle \dX,
     \Chi_{C}(\cH_L,\cH_R)\dX \rrangle  \quad \text{for all Borel sets $C\subset\mathbb{R}^2$}.
 \eeq
It follows from \eq{LF2}  that for each $T\ge 0$ and $\mu \in \RR$  
 the measure  
$ \Gamma_{\mu}^{T}$  can be written as
 \beq \label{GammaPhi}
 \Gamma_{\mu}^{T}(B)=\pi  \int_{\RR^2} \!\Phi(\d\lambda_{1}\d\lambda_{2}) \,
 F_\mu^T(\lambda_1,\lambda_2) \Chi_B(\lambda_1 - \lambda_2).
 \eeq
\end{remark}

We are thus led to the following definition.

\begin{definition}\label{defSigma2} The
 \emph{(ac-)conductivity measure} ($x_{1}$-$x_{1}$ component) at  $T=0$ and  $\mu\in \RR$
  is the finite positive  even Borel measure $ \Sigma_{\mu}^{0}$ on the real line given
   by
  \begin{equation}
     \label{kubo-expr2}
     \Sigma_{\mu}^{0} :=\psi(\mu)\bdelta_{0} +  \Gamma_{\mu}^{0}.
\end{equation}
The corresponding  in-phase linear-response current  is defined by
\begin{equation}\label{JinDef2}
  J_{\mathrm{lin}}^{\mathrm{in}}(t; 
  \mu,0,\cE):=   
  \int_{\mathbb{R}}\!\Sigma_{\mu}^{0}(\d\nu)\; \e^{\i\nu t}
  \widehat{\mathcal{E}}(\nu).
  \end{equation}
   \end{definition}
   
  \begin{remark}In view of Theorem~\ref{SigmaGamma} and  Remark~\ref{remfTlim}, 
   Definition~\ref{defSigma2}  agrees with  Definition~\ref{defSigma} on the common domain of definition, i.e., we have a unique definition for $\Sigma^0_\mu$ for all $\mu \in \RR$. 
  \end{remark} 
  
  \begin{remark} \label{remdeltafunctionfree}  In the absence of randomness, i.e., $H= - \Delta$, we may still carry out the above procedure and define $ \Sigma_{\mu}^{0}$ by \eq{kubo-expr2}
  with $\Psi$ as in \eq{defPsi} and  $\Gamma_{\mu}^{0}$ as in \eq{defGamma1} .
  In this case $\dX$ commutes with $H$, and hence $\cQ_{0} \dX= \dX$.  Thus
  $\Gamma_{\mu}^{0}=0$ and, for a Borel set $B\subset\mathbb{R}$,
\beq  
     \label{defPsi0}
     \Psi(B) =\pi  \llangle \dX,      \,\Chi_{B}(-\Delta)\dX\rrangle =\pi  \langle \pa[big]{ \delta_{\widehat{x}_{1}} -
     \delta_{-\widehat{x}_{1}}},    \Chi_{B}(-\Delta)\pa[big]{ \delta_{\widehat{x}_{1}} -
     \delta_{-\widehat{x}_{1}}}\rangle.
   \eeq  
It follows that $\Psi$ has a density given by  a continuous function $\psi$, the limit in \eq{limLDT} holds for every $\mu$, and (recall $\sigma(-\Delta)= [-2d,2d]$) 
\beq \label{nonzerodelta}
\Sigma_{\mu}^{0}= \psi(\mu)\bdelta_{0} \quad \text{with} \quad \psi(\mu)\begin{cases} > 0 & \text{if $\mu \in ]-2d,2d[$}\\
= 0 & \text{otherwise}
\end{cases}.
\eeq
Since  the {in-phase}
 conductivity $ \Re \sigma_{\mu}^0(\nu)$ is formally
the density  of $\Sigma_{\mu}^{0}$, \eq{nonzerodelta} is formally equivalent to the usual statement that for $H=-\Delta$ we have
\beq \label{nonzerodelta1}
\Re \sigma_{\mu}^0(\nu)= \psi(\mu) \delta(\nu),
\eeq
with $\delta(\nu)$ the formal Dirac delta function.
 \end{remark}
 
\begin{remark} 
  The picture described in Remark~\ref{remdeltafunctionfree} changes
  in the presence of any amount of randomness.  Let us introduce a
  disorder parameter in the Anderson Hamiltonian by setting
  $H_\omega^{(\lambda)}:= -\Delta + \lambda V_\omega$, where $\lambda
  \in \RR$ is the disorder parameter. Although the velocity operator
  $\dX$ does not depend on $\lambda$, any amount of randomness (i.e.,
  $\lambda \not=0$) implies $\cQ_{0}^{(\lambda)} \dX\not= \dX$ since
  then $[\dX, H_{\omega}^{(\lambda)}]= \lambda [\dX, V_{\omega}] \not=
  0$ for a.e.\ $\omega$. In the region of complete localization we
  know $\psi^{(\lambda)} (\mu)=0$ by Theorem~\ref{main0i}, and hence
  the conductivity measure has no atom at $0$ and we have
  \eq{SigmaGamma0}. At high disorder it is known that the region of
  complete localization (we include the complement of the spectrum) is
  the whole real line, in which case we can conclude that
  $\cQ_{0}^{(\lambda)} \dX=0$, i.e., $\cQ_{\perp}^{(\lambda)}
  \dX=\dX$.
  
  What happens if the Fermi energy $\mu$ lies in a spectral region
  where extended states are believed to exist is an open question.
  Common belief says that the conductivity is is nonzero in the region
  of extended states, but it is finite for all Fermi energies. The
  latter seems to rule out the existence of an atom of
  $\Sigma_{\mu}^{0}$ at $0$ for all Fermi energies, which is
  equivalent to having $\cQ_{0}^{(\lambda)} \dX=0$.  That would mean
  that any amount of disorder would have a very strong effect on the
  kernel of the Liouvillian, since we would have
  $\cQ_{\perp}^{(\lambda)} \dX=\dX$ for all $\lambda \not=0$ although
  we know that $\cQ_{0}^{(0)} \dX=\dX$.
\end{remark}

The justification for Definition~\ref{defSigma2} is given in the
following theorem.

\begin{theorem}\label{main1}
  \begin{nummer}
  \item 
    For all $T \ge 0$ the map $\mu \in \RR \mapsto \Sigma_\mu^T
    \in \cM_+^{(\e)}(\RR)$ is strongly measurable, and for every $T>0$ and
    $\mu \in \RR$ we have 
    \beq
      \begin{split}\label{convolution}
        \Sigma_{\mu}^{T}& = \pa[Big]{ \pa[normal]{-f_{0}^{T}}^\prime
          \ast \Sigma_{\bullet}^0}(\mu) , \quad\text{that is},\\ 
        \Sigma_{\mu}^{T}(B) & = \int_{\RR}\!\d E \; (-f_{\mu}^{T})'(E)
        \; \Sigma_{E}^{0}(B) \quad \text{for all Borel sets
          $B\subset\mathbb{R}$}.
      \end{split}
    \eeq 
  \item\label{main1ii}
    We have
    \beq
      \Sigma_{\mu}^{0}= 
      \begin{cases} 
        \slim_{T \downarrow 0}   \Sigma_{\mu}^{T}&  \text{for all  $\mu
          \in\Xi_0$} \\ 
        \wlim_{T \downarrow 0}   \Sigma_{\mu}^{T} & \text{for
          a.e. $\mu \in \RR \setminus \Xi_0$}  
      \end{cases} .
    \eeq 
  \item 
    We have 
    \beq \label{Jin01}
      J_{\mathrm{lin}}^{\mathrm{in}}(t; 
      \mu,0,\cE)=   \lim_{T \downarrow 0}
      J_{\mathrm{lin}}^{\mathrm{in}}(t;  
      \mu,T,\cE) \quad 
      \begin{cases}  \text{for all  $\mu \in\Xi_0$ }\\
        \text{for a.e.  $\mu \in \RR \setminus \Xi_0$}
      \end{cases} .
    \eeq 
  \end{nummer}
\end{theorem}

\section{Proofs}\label{secProofs}

In this section we prove Theorems~\ref{thmSigma}, \ref{main0} and
\ref{main1}.  We refer to Appendix~\ref{sec:Kubo} for the mathematical
framework and basic notation.

We start with a consequence of the Wegner inequality \cite{Weg81}.

\begin{lemma} \label{lemHLabscts} 
  $\cH_L$ and $\cH_R$ have purely absolutely continuous spectrum.
\end{lemma}
  
  \begin{proof} In view of \eq{JHJ} it suffices to prove that $\cH_L$ has purely absolutely continuous spectrum.  
  Given $\cK_2$, let $\eta_A\in\cM_+(\RR)$ be defined by
 \beq
\eta_A(B):=  \llangle A,
     \Chi_{B}(\cH_L)A \rrangle  \quad \text{for all Borel sets $B\subset\mathbb{R}$}.
 \eeq
 Since $\cK_\infty$ is dense in $\cK_2$, to prove the lemma it suffices to show that $\eta_A$ is absolutely continuous for all $A \in \cK_\infty$.  In this case, using \eq{reverse} and \eq{defN}, we get
 \beq \label{etaAB}
 \eta_A(B) = \tnorm{ \Chi_{B}(H)A }_2^2=  \tnorm{A^* \Chi_{B}(H) }_2^2
 \le \tnorm{A}_\infty^2 \tnorm{ \Chi_{B}(H) }_2^2 = \tnorm{A}_\infty^2 \cN(B).
 \eeq 
Since $\cN$ is absolutely continuous, we conclude that $\eta_A$ is also absolutely continuous. 
\end{proof}

\begin{lemma}\label{lemHS}
  For all  $g\in\mathcal{S}(\RR)$ we have
  \begin{equation}
    \cQ_0   [X_{1}, g(H)] =  \i g^\prime(\cH_L)   \cQ_0 \dX .
  \end{equation}
\end{lemma}

\begin{proof}
  The lemma is proved by means of the Helffer-Sj\"ostrand formula for
  smooth functions of self-adjoint operators (cf.
  \cite[Appendix B]{HuSi00}).  If $g \in\mathcal{S}(\RR)$, then for
  any self-adjoint operator $K$ we have
  \begin{align}\label{HS}
    g (K) &= \int_{\RR^{2}} \!\d\tilde{g}(z) \, (K-z)^{-1} ,\\
    g^\prime(K) &= - \int_{\RR^{2}} \!\d\tilde{g}(z) \, (K-z)^{-2} ,
  \end{align}
  where the integrals converge absolutely in operator norm.  Here $z=
  x + \i y$, $\tilde{g}(z)$ is an \emph{almost analytic extension} of
  $g$ to the complex plane, and $\d\tilde{g}(z) := \frac 1
  {2\pi}\partial_{\bar{z}}\tilde{g}(z) \,\mathrm{d} x\, \mathrm{d} y $
  with $\partial_{\bar{z}}= \partial_x + \i
  \partial_y$.

  Thus, for $g\in\mathcal{S}(\RR)$ we have, with $R_\omega (z)=
  (H_\omega-z)^{-1}$, $\cR_L(z)= (\cH_L -z)^{-1}$, $\cR_R(z)= (\cH_R
  -z)^{-1}$,
  \begin{align}
    [X_{1}, g(H)] & =   \int_{\RR^{2}} \!\d\tilde{g}(z) \, [X_1, R(z)]
    = - \i \int_{\RR^{2}} \!\d\tilde{g}(z) \, 
    R(z) \dX R(z) \notag \\
    &= - \i  \int_{\RR^{2}} \!\d\tilde{g}(z) \, \cR_L(z)\cR_R(z) \dX .  
  \end{align}
  We recall $ [X_{1}, g(H)], [X_1, R(z)] \in \cK_2$, and the integrals
  converge absolutely in operator norm in $\cK_2$ (see
  \cite[Proposition~2.4]{BGKS} and its proof).  It follows, using
  \eq{QH=QH}, that
  \begin{equation}
    \cQ_0 [X_{1}, g(H)] = - \i  \int_{\RR^{2}}
    \!\d\tilde{g}(z) \, \cR_L(z)^2 \cQ_0 \dX=  \i g^\prime(\cH_L)
    \cQ_0 \dX . 
  \end{equation}
\end{proof}

The following lemma plays an important role in our analysis.

\begin{lemma}
  \label{Y-rep}
  \begin{nummer}
  \item 
    If either $T>0$ or $\mu \in \Xi_0 $, we have
    \beq \label{YLX}
      \cF_\mu^T \dX= - \cL Y_\mu^T .
    \eeq
    In particular, we conclude that $ \cF_\mu^T \dX \in \cD
    \pa[normal]{\cL_\perp^{-1}}$. 
  \item  
    Let   $T>0$. Then for  all  $\mu\in\RR$  we have   
    \begin{equation}\label{Ydecomp}
      Y_{\mu}^{T} =\pa[normal]{ -f_{\mu}^{T}}^\prime (\cH_L)
      \;\cQ_{0}    \dX - \cL_\perp^{-1} \cF_\mu^T \dX . 
    \end{equation}
  \end{nummer}
\end{lemma}

\begin{proof} 
  Let either $T>0$ or $\mu \in \Xi_0 $, so $Y_\mu^T \in \cK_2$.  Given
  $\varphi\in\ell^{2}(\mathbb{Z}^{d})$ with compact support, we have
  \begin{align}
    \cF_\mu^T \dX  \varphi
    &  = \i \Bigl\{ f_{\mu}^{T}(H)  [H, X_{1}] -
    [H,X_{1}] 
    f_{\mu}^{T}(H) \Bigr\} \varphi \nonumber\\
    &   = -\i \Bigl\{ H  [X_{1},  f_{\mu}^{T}(H) ] -
    [X_{1},  f_{\mu}^{T}(H) ]  H \Bigr\}\varphi  \nonumber\\
    &=  -( \cH_{L} -\cH_{R}) Y_{\mu}^{T} \varphi=- \cL Y_{\mu}^{T} \vphi,    
  \end{align}
  since $ f_{\mu}^{T}(H) \phi \in \cD(X_1)$ for
  $\phi\in\ell^{2}(\mathbb{Z}^{d})$ with compact support by
  \eq{assumpIntro}.  Thus \eq{YLX} follows, and, in view of \eq{FQ},
  we have $ \cF_\mu^T \dX \in \cD\pa[normal]{\cL_\perp^{-1}}$.
  
  We now let $T>0$, and note that \eq{Ydecomp} follows from \eq{YLX}
  since Lemma~\ref{lemHS} gives 
  \beq 
    \cQ_0 Y_{\mu}^{T}= \pa[normal]
    {- f_{\mu}^{T}}^\prime (\cH_L) \, \cQ_{0} \dX . 
  \eeq
\end{proof}

\begin{lemma}\label{lemYTcnt} The
     map $]0,\infty[ \ni T \mapsto Y _{\mu}^{T} \in \cK_2$ is  
norm continuous  for every $\mu\in\RR$.
\end{lemma}

\begin{proof} If $g\in\mathcal{S}(\RR)$, it follows from  \cite[Proposition~2.4]{BGKS} and \eq{norminfinity2} that
\beq
\tnorm{[X_1, g(H)]}_2 \le \tnorm{[X_1, g(H)]}_\infty \le C \hnorm{g}_3, 
\eeq
where $C$ is a constant depending only on $H$ and
\begin{equation} \label{sdfn}
  \hnorm{g}_3 := \sum_{r=0}^3 \int_{\mathbb{R}}\!\mathrm{d}u\;
  |g^{(r)}(u)|\,(1 + \abs{u}^{2})^{\frac {r-1} 2}  . 
\end{equation}
The lemma follows in view of \eq{defY}.
\end{proof}

We are ready to prove  Theorem~\ref{thmSigma}.   Note that for all
$T \ge 0$ and $\mu \in \RR$ we have 
\beq
  0 \le \pa[big]{ \cF_\mu^T}^2 \le 1.   \label{F21}
\eeq
Moreover, for all $\mu \in \RR$ the operator $ \pa[big]{ \cF_\mu^0 }^2$ is an orthogonal projection in $\cK_2$,   and  hence
\beq  \label{F03}
  \pa[big]{ \cF_\mu^0 }^3=  \cF_\mu^0 .
  \eeq
In addition, if $\mu \in \Xi_0$  we have 
\begin{gather}
\label{0Tswitch0}
  \pa[big]{ \cF_\mu^0 }^2 Y_\mu^0 = Y_\mu^0 ,\\   \cF_\mu^0 Y_\mu^T  =  \cF_\mu^T Y_\mu^0
  \quad \text{for all $T \ge 0$}. \label{0Tswitch}
  \end{gather}

\begin{proof}[Proof of Theorem~\ref{thmSigma}]
Let    $\mu \in \Xi_0 $ and $\Sigma_\mu^0$ be given by \eq{kubo-expr}.
Using \eq{0Tswitch0} and \eq{YLX}, we have 
\begin{align}
 \Sigma_{\mu}^{0}(B) &=  \pi \llangle \dot{X}_{1},
     \Chi_{B}(\mathcal{L})\pa[big]{ \cF_\mu^0 }^2  Y_{\mu}^{0} \rrangle = 
     \pi \llangle \cF_\mu^0   \dot{X}_{1},
     \Chi_{B}(\mathcal{L})\cF_\mu^0  Y_{\mu}^{0} \rrangle \notag\\
 &   =  \pi \llangle Y_\mu^0  ,
     \Chi_{B}(\mathcal{L}) (-\cL)\cF_\mu^0  Y_{\mu}^{0} \rrangle,
\end{align}
 and hence coincides with  \cite[Eq. (3.31)]{KLM}, a finite positive  even  Borel measure by  \cite[Theorem~3.4]{KLM}.
 
If $T>0$ and $\mu \in \RR$ arbitrary, we use \eq{Ydecomp} to rewrite
$\Sigma_\mu^T$ given by \eq{kubo-expr} as in \eq{Sigmadecomp}, where
$\Psi$, given by \eq{defPsi}, is clearly in $\cM_+(\RR)$, and $
\Gamma_{\mu}^{T}$, given in \eq{defGamma12}, is also seen to be in
$\cM_+(\RR)$ by \eq{LF2}.  We conclude that $\Sigma_\mu^T \in
\cM_+(\RR)$.  The same argument as in \cite[Proof of
Theorem~3.4]{KLM} shows that the measure $ \Gamma_{\mu}^{T}$, and
hence also $\Sigma_\mu^T$, is even.
 
To prove \eq{SigmaR}, note that for either $T>0$ or $\mu \in \Xi_0$ it
follows from \eq{kubo-expr}, the Cauchy--Schwarz inequality and
$|f_{\mu}^{T}| \le 1$, that
\begin{align}
  \Sigma_{\mu}^{T}(\RR) & = - \pi\mathbb{E} \bigl\{ \langle X_ 
  {1}^{2} H_\omega
  \delta_{0}, f_{\mu}^{T}(H_\omega)\delta_{0}\rangle \bigr\} =
  - \pi  \mathbb{E} \bigl\{ \langle \delta_{\widehat{x}_{1}} +
  \delta_{-\widehat{x}_{1}}, f_{\mu}^{T}(H_\omega)\delta_{0}\rangle \bigr 
  \} \notag\\
  &  \le   \sqrt{2} \, \pi \tnorm[big]{f_{\mu}^{T}(H)}_2  \le \sqrt{2}\,
  \pi \tnorm[big]{f_{\mu}^{T}(H)}_\infty \le  \sqrt{2} \, \pi  . 
\end{align}

 We have thus proved parts (i) and (ii). Part (iii) is an immediate consequence of Lemma~\ref{lemYTcnt}. To prove (iv), 
 given a bounded measurable function $g$ and $T\ge0 $, we write
  \begin{equation} 
    \label{decompose}
    \Sigma_{\mu}^{T}(g)= \pi \llangle \dot{X}_{1}, g(\mathcal{L})
    (\cF_\mu^0)^2 Y_{\mu}^{T} \rrangle + \pi \llangle \dot{X}_{1},
    g(\mathcal{L}) \bigl(1 -(\cF_\mu^0)^2\bigr) Y_{\mu}^{T} \rrangle.
  \end{equation}
  In view of \eq{F21}, the same argument used to prove $\Sigma_\mu^T \in \cM_+(\RR)$ shows that  both terms on the right-hand side of \eqref{decompose} 
  are integrals of $g$ with respect to finite
  positive Borel measures on $\RR$. On account of \eqref{0Tswitch} we have
  \begin{equation}
    \llangle \dot{X}_{1},
    g(\mathcal{L})  (\cF_\mu^0)^2 Y_{\mu}^{T} \rrangle = 
    \llangle  \dot{X}_{1}, 
    g(\mathcal{L}) \cF_\mu^0 \cF_\mu^T Y_{\mu}^{0} \rrangle=  \llangle \cF_\mu^T  \dot{X}_{1}, 
    g(\mathcal{L}) \cF_\mu^0 Y_{\mu}^{0} \rrangle.
  \end{equation}
Using the Cauchy--Schwarz inequality, we get
\beq  \tnorm[big]{(\cF_{\mu}^{T} - \cF_{\mu}^{0})\dot{X}_{1}}_{2} \le 2 \tnorm[big]{\dX}_\infty  \tnorm[big]{f_{\mu}^{T}(H) - f_{\mu}^{0}(H)}_{2} .
\eeq
Recalling \eq{defN}, we have
\beq
\tnorm[big]{f_{\mu}^{T}(H) - f_{\mu}^{0}(H)}_{2}^2=\int_{\RR}\!\cN(\d
E) \; \bigl| f_{\mu}^{T}(E) - f_{\mu}^{0}(E) \bigr|^2,
\eeq
and hence
\beq  \label{f-fto0}
\lim_{T\downarrow 0} \tnorm[big]{f_{\mu}^{T}(H) - f_{\mu}^{0}(H)}_{2} =0
\eeq
by dominated convergence. It follows that
 $\lim_{T\downarrow 0}
  \tnorm[big]{(\cF_{\mu}^{T} - \cF_{\mu}^{0})\dot{X}_{1}}_{2} =0$.
 We conclude, using \eq{0Tswitch0}, that
  \begin{equation}
    \label{oneterm}
    \pi \lim_{T\downarrow 0} \llangle \dot{X}_{1},
    g(\mathcal{L})  (\cF_\mu^0)^2 Y_{\mu}^{T} \rrangle 
    =  \pi  \llangle  \cF_\mu^0 \dot{X}_{1}, 
    g(\mathcal{L}) \cF_\mu^0  Y_{\mu}^{0} \rrangle=
    \Sigma_{\mu}^{0}(g).
  \end{equation}
  On the other hand, it follows from \eq{SigmaR} that
  \begin{equation}
    \lim_{T\downarrow 0}  \Sigma_{\mu}^{T}(\RR)=  \Sigma_{\mu}^{0}(\RR).
  \end{equation}
  Combining this with \eqref{oneterm}, where we set $g=1$, we conclude that
  \begin{equation}
    \lim_{T\downarrow 0} \llangle \dot{X}_{1},
    \bigl( 1 -(\cF_\mu^0)^2\bigr) Y_{\mu}^{T} \rrangle =0.
  \end{equation}
Since $\llangle \dot{X}_{1}, \Chi_B(\mathcal{L})
   \bigl( 1 - (\cF_\mu^0)^2 \bigr) Y_{\mu}^{T} \rrangle$ is a positive
  measure, it converges to $0$ strongly.  Part (iv) is proven.
  
  It remains to prove part (v).  Let  $\mu\in\Xicl$, so $Y_{\mu}^{T}\in \cK_2$ for all $T\ge 0$.
  We need to prove that 
   \begin{equation}\label{YTconv}
     \lim_{T\downarrow 0} \;\tnorm[big]{Y_{\mu}^{T} - Y_{\mu}^{0}}_ 
{2} =0.
   \end{equation}
  Standard calculations give
    \begin{align}
     \label{Y-bound}
     \tnorm[big]{Y_{\mu}^{T} - Y_{\mu}^{0}}_{2}^{2}  & =
     \mathbb{E} \Bigl\{ \Big\langle  \bigl(f_{\mu}^{T}(H) -
     f_{\mu}^{0}(H)\bigr) \delta_{0}, X_{1}^{2} \bigl(f_{\mu}^{T}(H) -
     f_{\mu}^{0}(H)\bigr) \delta_{0} \Big\rangle \Bigr\}  \nonumber \\
     & \le  \;\tnorm[big]{f_{\mu}^{T}(H) - f_{\mu}^{0}(H)}_{2} \;
     \pa{\mathbb{E} \Bigl\{ \norm[big]{ X_{1}^{2} \bigl(f_{\mu}^{T}(H) -
     f_{\mu}^{0}(H)\bigr) \delta_{0} }^2 \Bigr\}}^{\frac 1 2} .
   \end{align}
  In view of \eq{f-fto0}, the desired \eq{YTconv} follows if we prove that
  \beq  \label{limsupff}
  \limsup_{T\downarrow 0}  \mathbb{E} \Bigl\{ \norm[big]{ X_{1}^{2} \bigl(f_{\mu}^{T}(H) -
     f_{\mu}^{0}(H)\bigr) \delta_{0} }^2 \Bigr\} < \infty .
  \eeq
  
  To prove \eq{limsupff} we use that $\mu \in \Xicl$, and hence there exists $\delta>0$ such that   ${I_\delta} \subset \Xicl$, where $I_\eta:= ]\mu - \eta, \mu + \eta[$ for $\eta>0$.  We pick functions $g_j \in C_{\mathrm{c}}^\infty(\RR)$, $j=1,2$, such that  $0\le g_j \le 1$,
  $ \Chi_\mathfrak{S}=  (g_1 + g_2) \Chi_\mathfrak{S}$, $\supp g_1 \subset I_\delta$,
$\supp g_2 \subset \RR \setminus I_{\frac \delta 2}$.  Letting  $g_{\mu}^{T} = f_{\mu}^{T}-f_{\mu}^{0}$, we have  
\beq
f_{\mu}^{T}(H) - f_{\mu}^{0}(H)= g_{\mu}^{T}(H)=g_{\mu}^{T}(H)g_1(H) + g_{\mu}^{T}(H)g_2(H).
\eeq

Since  $\supp g_1 \subset \Xicl$ and $\big| g_{\mu}^{T} \big| \le 2$ for all $T > 0$, standard estimates \cite{Aiz94, AiGr98,GKcmp,GKjsp} give
\beq
\sup_{T>0} \mathbb{E} \Bigl\{ \norm[big]{ X_{1}^{2}g_{\mu}^{T}(H)g_1(H) \delta_{0} }^2 \Bigr\} < \infty.
\eeq
On the other hand, explicit calculations show that 
\beq
\sup_{T>0} \norm[Big]{ \pa[big]{g_{\mu}^{T}}^{(k)}\Chi_{\RR \setminus I_{\frac \delta 2}}}_\infty < \infty \quad \text{for all $k=0,1,2\ldots$}.
\eeq
Since $\supp g_2 \subset \RR \setminus I_{\frac \delta 2}$, a calculation using \cite[Theorem~2]{GKpams} shows that
\beq
\sup_{T>0} \mathbb{E} \Bigl\{ \norm[big]{ X_{1}^{2}g_{\mu}^{T}(H)g_2(H) \delta_{0} }^2 \Bigr\} < \infty.
\eeq
The estimate\eq{limsupff} follows. 
\end{proof}

We now turn to  Theorem~\ref{main0}.

\begin{proof}[Proof of Theorem~\ref{main0}]
  Note that we already proved parts (iii) and (iv) while proving
  Theorem~\ref{thmSigma}. To prove (v), note that it follows from
  \eq{kubo-expr}, \eq{0Tswitch0}, \eq{HQL}, \eq{YLX}, and \eq{F03}
  that for all Borel sets $B\subset\mathbb{R}$ we have
  \begin{align} 
    \label{decompose2}
    \Sigma_{\mu}^{0}(B)&= \pi \llangle \dot{X}_{1}, \Chi_B(\cL)
    (\cF_\mu^0)^2 Y_{\mu}^{0} \rrangle =  \pi \llangle \dot{X}_{1}, \Chi_B(\cL)
    (\cF_\mu^0)^2    \cL_\perp^{-1}\cL  Y_{\mu}^{0} \rrangle \notag \\
    &= - \pi \llangle \dot{X}_{1}, \Chi_B(\cL)
    (\cF_\mu^0)^2    \cL_\perp^{-1}\cF_{\mu}^{0}\dX \rrangle= \Gamma_\mu^0(B) .
  \end{align}

  Now, we turn to part~\itemref{main0i}. Let $\Psi$ be given by
  \eq{defPsi}, it is clearly in $ \cM_+(\RR)$. Since
  \beq
    \dX \delta_0 = - \i \pa[big]{ \delta_{\widehat{x}_{1}} -
      \delta_{-\widehat{x}_{1}}}, 
  \eeq
  we have,  for all Borel sets $B\subset\mathbb{R}$,  recalling \eq{defN},
  \begin{align}
    \tfrac 1 \pi \Psi(B) &\le \llangle \dX, \Chi_{B}(\cH_L)\dX\rrangle
    = \EE\set{ \langle \pa[big]{ \delta_{\widehat{x}_{1}} -
        \delta_{-\widehat{x}_{1}}}, \Chi_{B}(H)\pa[big]{
        \delta_{\widehat{x}_{1}} -
        \delta_{-\widehat{x}_{1}}}\rangle}    \notag\\
    & \le 2 \cN(B) + 2 \, \EE\set{\norm[big]{
        \Chi_{B}(H)\delta_{\widehat{x}_{1}}} \norm[big]{
        \Chi_{B}(H)\delta_{-\widehat{x}_{1}}} } \le 4\cN(B).
  \end{align}
  It follows that $\Psi$ is absolutely continuous with respect to the
  density of states measure $\cN$, and that its density with respect
  to Lebesgue measure, $\psi$, satisfies $\psi(E) \le 4\pi n(E)$ for
  Lebesgue-a.e.\ $E\in\RR$. 
  Since the functions
  $\pa[normal]{-f_{0}^{T}}^{\pr}$ form an approximate identity as $T
  \downarrow 0$, it follows from the absolute continuity of $\Psi$ and
  the Lebesgue Differentiation Theorem
  (cf. \cite[Corollary~2.1.17]{Gra}) that  
  \beq
    \label{limLDT} 
    \lim_{T \downarrow 0} \Psi\pa[big]{\pa[normal]{-
        f_{\mu}^{T}}^{\pr}}= \psi(\mu) \quad \text{for a.e.\ $\mu$}.
  \eeq 
  From parts \itemref{sameKLM} and \itemref{slimSig} of Theorem~\ref{thmSigma}
  and \eq{Sigma0} (which is proved already) we conclude that $\lim_{T
    \downarrow 0} \Psi\pa[big]{\pa[normal]{- f_{\mu}^{T}}^{\pr}}=0$
  for Lebesgue-almost all $\mu \in \Xi_0$.
  Theorem~\ref{main0i} is proven.
 
  To finish, we need to prove part (ii).  Let $\Phi \in \cM_+(\RR^2)$
  be the velocity-velocity correlation measure given  in \eq{vel-vel}.     
  As a consequence of \eq{GammaPhi}, \eq{Sigmadecomp} and \eq{SigmaR},
  we have 
  \beq\label{intPhibound}
    \int_{\RR^2}  \!\Phi(\d\lambda_{1}\d\lambda_{2}) \,
    F_\mu^T(\lambda_1,\lambda_2)\le \sqrt{2} \quad \text{for all $T>0$
      and $\mu \in \RR$}. 
  \eeq
But for all $\mu \in \RR$ we have
\beq
\lim_{T \downarrow 0}  F_\mu^T(\lambda_1,\lambda_2)= F_\mu^0(\lambda_1,\lambda_2)
\quad \text{for $\Phi$-a.e.\ $(\lambda_1,\lambda_2)\in \RR^2$},
\eeq 
where we used the fact that the two marginals of $\Phi$ are absolutely continuous, a consequence of Lemma~\ref{lemHLabscts}.  (More is true: the two marginals are equal to the measure $\eta_{\dX}$, and hence have a bounded  density, cf. \eq{etaAB}.)  Using Fatou's Lemma and \eq{intPhibound} we conclude that for all $\mu \in \RR$ we have
\beq
\int_{\RR^2}  \!\Phi(\d\lambda_{1}\d\lambda_{2}) \,
 F_\mu^0(\lambda_1,\lambda_2)\le \liminf_{T \downarrow 0} \int_{\RR^2} \!\Phi(\d\lambda_{1}\d\lambda_{2}) \,
 F_\mu^T(\lambda_1,\lambda_2)\le \sqrt{2}.
\eeq
 Theorem~\ref{finiteGamma}  follows.
 \end{proof}

It remains to prove Theorem~\ref{main1}.

\begin{proof}[Proof of  Theorem~\ref{main1}]
  To prove part (i), we remark that measurability in $\mu$ follows
  from \eqref{Sigmadecomp} and \eqref{GammaPhi} if $T>0$, respectively
  from Definition~\ref{defSigma2} and \eqref{GammaPhi} if $T=0$. 
  Now, Definition~\ref{defSigma2}, Theorem~\ref{main03}, and
  Theorem~\ref{main0i} imply that it suffices to prove \eq{convolution} with
  $\Gamma_\mu^T$ substituted for $\Sigma_\mu^T$, that is,
  \beq\label{convolution2}
    \Gamma_{\mu}^{T}(B)  =  \int_{\RR}\!\d E \; {(-f_{\mu}^{T})'(E)}
    \;\Gamma_{E}^{0}(B) \quad \text{for all Borel sets $B\subset\mathbb{R}$}.
  \eeq
  But this follows from \eq{GammaPhi} using Fubini's Theorem plus the fact that
  \beq
    f_{\mu}^{T}(t)=   \int_{\RR}\!\d s \; {(-f_{\mu}^{T})'(s)} \;
    f_{s}^{0}(t) \quad \text{for all $t \in \RR$}. 
  \eeq

  Next we turn to part (ii). As in the proof of \eqref{limLDT}, it
  follows from \eq{convolution} and the Lebesgue Differentiation
  Theorem that for each Borel set $B \subset \RR$ we have $\lim_{T
    \downarrow 0} \Gamma_{\mu}^{T}(B) = \Gamma_{\mu}^{0}(B) $ for
  Lebesgue-a.e.\ $\mu\in\RR$ (the exceptional set depending on $B$!).
  Let $\set{I_n}_{n \in \NN}$ denote an enumeration of 
  the bounded intervals with rational endpoints.  It follows that for
  a.e.\ $\mu$ we have $\lim_{T \downarrow 0} \Gamma_{\mu}^{T}(I_n) =
  \Gamma_{\mu}^{0}(I_n) $ for all $n \in \NN$, and hence we have
  $\wlim_{T \downarrow 0} \Gamma_{\mu}^{T}= \Gamma_{\mu}^{0}$ for
  a.e.\ $\mu$.  Part (ii) now follows using Theorem~\ref{slimSig} for
  $\mu \in \Xi_0$.

  Part (iii) is an immediate consequence of part (ii).
\end{proof}

\appendix

\section{The mathematical framework for linear response theory}
\label{sec:Kubo}
In this appendix we recall the mathematical framework for linear response
theory, following \cite[Section~3]{BGKS} and \cite[Section~3]{KLM}
(see also \cite{BESB,SBB}).  We restrict ourselves to the Anderson
model.  The Hamiltonian $H_{\omega}$, given in \eq{AND}, is a
measurable map from the probability space $(\Omega, \PP)$ to the
bounded self-adjoint operators on $\cH=\ell^2(\ZZ^d)$. The probability
space $(\Omega, \PP)$ is equipped with an ergodic group $\{\tau_{a}; \
a \in \ZZ^d\}$ of measure preserving transformations, satisfying the
covariance relation
\begin{align}
 \label{covintro}
 U(a) H_\omega U(a)^* = H_{\tau_{a}(\omega)}  \quad \text{for all
   $a \in \ZZ^{d}$}, 
\end{align}
where $U(a)$ denotes translation by $a$, i.e., $U(a) \delta_b
:=\delta_{b+a}$ when applied to any member of the canonical orthonormal
basis $\{\delta_{b}; b \in \ZZ^{d}\}$ for $\ell^2(\ZZ^d)$.

Let $\cH_{c}=\ell_c^2(\ZZ^d)$ be the (dense) subspace of finite linear combinations of
the canonical basis vectors. By $\cK_{mc}$ we denote the vector space
of measurable covariant operators $A\colon \Omega \to
\mathrm{Lin}\bigl(\cH_{c}, \cH)$, identifying measurable covariant
operators that agree $\PP$-a.e.; all properties stated are assumed to
hold for $\PP$-a.e.\ $\omega \in \Omega$.  Here
$\mathrm{Lin}\bigl(\cH_{c}, \cH)$ is the vector space of linear
operators from $ \cH_{c}$ to $ \cH$.  Recall that $A$ is measurable if
the functions $\omega \to \langle \phi, A_{\omega} \phi\rangle$ are
measurable for all $\phi \in \cH_{c}$, $A$ is covariant if
\begin{equation}\label{covariant}
  U(a)A_{\omega} U(a)^{*}= A_{\tau_{a}(\omega)} \quad \text{for all} \quad a
  \in \ZZ^{d}.
\end{equation}
It follows (for $\cH=\ell^2(\ZZ^d)$) that  $\mathcal{D}(A_\omega^*) \supseteq \cH_{c}$ for
$A\in\cK_{mc}$, i.e., $A$ is locally bounded. 
Thus, the operator $A_\omega^\ddagger:=A_\omega^*{\big|}_{\cH_{c}}$ is
well defined. Note that $(\cJ A)_{\omega}:= A_\omega^\ddagger$ defines
a conjugation in $\cK_{mc}$.

We introduce norms on $\cK_{mc}$ given by
\begin{equation}
  \begin{split}
    \tnorm{{A}}_\infty &:=  \|\, \|{A_\omega} \| \,
    \|_{\mathrm{L}^{\infty}(\Omega, \PP)},\\ 
    \tnorm{{A}}_p^{p} &:= \E\bigl\{\langle \delta_{0},
    |\overline{A_{\omega}}|^{p}  
    \delta_{0} \rangle\bigr\}, \quad p=1,2,  \end{split}
\end{equation}
and consider the normed spaces
\begin{equation}
  \cK_{p}: = \{ {A}\in \cK_{mc}; \, \tnorm{{A}}_p<\infty \}, \quad
  p=1,2,\infty . 
\end{equation}
It turns out that $\cK_{\infty}$ is a Banach space and $\cK_{2}$ is a
Hilbert space with inner product
\begin{equation}
    \llangle A, B\rrangle :=  \ \EE \bigl\{\langle A_{\omega}\delta_{0}, B_{\omega} 
    \delta_{0}\rangle
    \bigr\}  ,
\end{equation}
and we have
\beq \label{reverse}
 \llangle A, B\rrangle =  \llangle B^\ddagger, A^\ddagger \rrangle
\eeq
Since $\cK_{1}$ is not complete, we introduce its (abstract)
completion $\overline{\cK_{1}}$. The conjugation $\cJ$ is an isometry
on each $\cK_{p}$, $ p=1,2,\infty $.  We also have
\begin{equation} \label{norminfinity2}
  \tnorm{{A}}_1 \le \tnorm{{A}}_2 \le  \tnorm{{A}}_\infty \quad \text{and
    hence} \quad  \cK_{\infty} \subset \cK_{2} \subset \cK_{1},
\end{equation}
and $ \cK_{\infty}$ is dense in $\cK_{p}$, $p=1,2$.  Moreover, we have
$H,\Delta,\dot{X}_{1}\in \cK_{\infty}$.

Given $A\in \cK_{\infty}$, we identify $A_{\omega}$ with its closure
$\overline{A_{\omega}}$, a bounded operator in $\cH$.  We may then
introduce a product in $ \cK_{\infty}$ by pointwise operator
multiplication, and $\cK_{\infty}$ becomes a $C^{*}$-algebra.
($\cK_{\infty}$ is actually a von Neumann algebra
\cite[Subsection~3.5]{BGKS}.)  This $C^{*}$-algebra acts by left and
right multiplication in $\cK_{p}$, $p=1,2$.  Given $A \in \cK_{p}$, $B
\in \cK_{\infty}$, left multiplication $B \omL A$ is simply
defined by $(B \omL A)_{\omega} := B_{\omega} A_{\omega}$.  Right
multiplication is more subtle, we set $(A \omR B)_{\omega} :=
A_{\omega}^{\ddagger*} B_{\omega}$ (see \cite[Lemma~3.4]{BGKS} for a
justification), and note that $(A \omR B)^{\ddagger}= B^{*}
\omL A^{\ddagger}$.  Moreover, left and right multiplication
commute:
\begin{equation}
  B \omL A \omR C:= B \omL ( A \omR C)=(B \omL A )\omR C 
\end{equation}
for $A \in \cK_{p}$, $B,C \in \cK_{\infty}$. We refer to
\cite[Section~3]{BGKS} for an extensive set of rules and properties
which facilitate calculations in these spaces of measurable covariant
operators.

Since $H \in \cK_{\infty}$, we define bounded commuting self-adjoint
operators $\cH_{L}$ and $\cH_{R}$ on $\cK_{2}$ by
\begin{equation}
  \cH_{L}A := H\omL A\quad \text{and} \quad  \cH_{R}A := A\omR H; 
\end{equation}
note that 
\begin{equation} \label{JHJ}
  \cH_{R} = \cJ \cH_{L} \cJ.  
\end{equation} 
The \emph{Liouvillian}
is then defined by
\begin{equation}\label{LHH}
  \cL := { \cH_{L}- \cH_{R}}  ,
\end{equation}
and hence satisfies
\begin{equation} \label{JLJ}
  \cL = - \cJ \cL \cJ.  
\end{equation} 
Note that (cf. \cite[argument below Eq.~(5.91)]{BGKS}) 
\beq
\label{kerL} \ker \cL= \set{A \in \cK_2; \; A\omL f(H)= f(H)\omR A \;
  \text{for all $f \in \cS(\RR)$}}.  
\eeq

The \emph{trace per unit volume} is given by
\begin{equation}\label{tuv1}
  \cT(A) :=\E \bigl\{\langle \delta_{0}, {A}_\omega \delta_{0} 
  \rangle\bigr\}    \quad\text{for}\quad 
  A \in \cK_1,
\end{equation}
a well defined linear functional on $\cK_1$ with $|\cT(A)| \le
\tnorm{A}_1$, and hence can be extended to $\overline{\cK_{1}}$.  Note
that $\cT$ is indeed the {trace per unit volume}:
\begin{equation} \label{tuv}
  \cT(A) = \lim_{L \to \infty}\, \textstyle{ \frac 1 {|\Lambda_L|}}
  \tr\left\{\Chi_{\Lambda_L} A_\omega \Chi_{\Lambda_L}\right\}
\quad \text{for $\PP$-a.e. $\omega$} \, , 
\end{equation}
where $\Lambda_L$ denotes the cube of side $L$ centered at $0$ (see
\cite[Proposition~3.20]{BGKS}).  Moreover,
\begin{equation}
    \llangle A, B\rrangle=  \cT \set{A^{*} B }
    \quad \text{for all  $A,B \in \cK_{2}$}.
\end{equation}

\section{The region of complete localization} \label{Xicl}

There is a wealth of localization results for the Anderson model in arbitrary
dimension, based either on the multiscale analysis \cite{FrSp83, FrMa85,
  DrKl89}, or on the fractional moment method \cite{AiMo93, Aiz94}.
The spectral region of applicability of both methods turns out to be the same,
and in fact it can be characterized by many equivalent conditions
\cite{GKduke,GKjsp}.  For this reason we call it the \emph{region of complete
  localization} as in \cite{GKjsp}.
  
   The most convenient definition for this
paper is by the conclusions of  \cite[Theorem~3]{GKjsp}. For convenience we include the complement of the spectrum in the region of complete
  localization.

\begin{definition}
  \label{locdef}
  The region of complete localization $\Xi^{\mathrm{cl}}$ for the Anderson
  Hamiltonian $H$ is the set of energies $E \in \RR$ for which there is an
  open interval $I \ni E$ and  constants $\zeta>0$ and $C< \infty$ such that
  \begin{equation} \label{fermidecay} 
    \mathbb{E}\left\{\sup_{\mu \in
        I}\abs{ \langle \delta_x, f^0_{\mu}\pa{H_{\omega}}
        \delta_{0}\rangle}^2\right\} \leq C \,\mathrm{e}^{-|x|^\zeta}
    \quad \text{for all $x \in \ZZ^d$}.
  \end{equation} 
\end{definition}

\begin{remark} 
  As remarked in the comments below \cite[Theorem~3]{GKjsp}, it
  suffices to require fast enough polynomial decay in \eq{fermidecay};
  subexponential decay then follows.
\end{remark}

\begin{remark}  For the Anderson model, it follows from \cite{Aiz94,AiGr98} that we have exponential decay in   \eq{fermidecay}.  More precisely, if $E \in \RR \in \Xi^{\mathrm{cl}}$,
 there is an open interval $I \ni E$ and  constants $m>0$ and $C< \infty$ such that
 \begin{equation} \label{fermidecay2}
\mathbb{E}\left\{\sup_{\mu \in I}\abs{ \langle \delta_x, f^0_{\mu}\pa{H_{\omega}}
\delta_{0}\rangle}^2\right\}
 \leq   C \,\mathrm{e}^{-m |x|} \quad \text{for all $x \in \ZZ^d$}.
\end{equation} 
\end{remark}


\end{document}